# Growth of metamaterial from Isolated nuclei with anisotropic building block

Kyoung Hwan Choi[1], Da Young Hwang[1] and Dong Hack Suh[1†]

*1 Advanced Materials & Chemical Engineering Building 311, 222 Wangsimni-ro, Seongdong-Gu, Seoul, Korea, E-mail: dhsuh@hanyang.ac.kr*

**Abstract**

Crystallization has long been the subject of research as one of the basic ways in which solid materials are constructed. In particular, the nucleation stage has not been isolated, thus has been predicted through many calculations and achieved theoretical completion through the nucleation rate(J). Since most of these results were obtained through isotropic building blocks in three-dimensional space, it was difficult to interpret nuclei formed by anisotropic building block in 2D or 1D structure. Recently, a lot of studies related to amyloid fibril have shown nucleation of anisotropic building block. However, due to the complexity of the amyloid fibrils, there is no unified explanation of the thermodynamic method of classical nucleation theory which is the energy loss from surface and energy gain from volume. We have experimentally demonstrated the isolation of nuclei of the orthorhombic phase of HYLION-12 which is a Dirac metamaterial and provide the effect of anisotropy of the molecules on nucleation The thermal behavior of nuclei of Dirac metamaterial through DSC has demonstrated that it can be crystallized to a Dirac metamaterial through the first order phase transition. The growth process is verified at low temperature where no phase transition occurs. The calculation of surface and bulk energy of the Dirac metamaterial was conducted. It could explain the isolation of nuclei of the Dirac metamaterial by enlarging the thermodynamic classical nucleation theory.

**Introduction**

Nucleation is the process of generation of crystalline clusters that can go into growth phase.[1] Unless the size of the smallest cluster exceeds the critical nuclei size, the clusters are likely to be dissolved.[2] Nuclei are generally regarded as a transition state, and cannot be isolated.[3] Therefore, isolation of nuclei has heretofore considered to be difficult in the relevant academia. The nucleation rate(J) which is defined as the number of supernuclei that form in a supersaturated old phase per unit volume is used to study nucleation and growth.[4,5] It has deeply related with important quantities such as the size distribution and the maximum number of crystals formed in the system, the rate of nucleation-mediated crystal growth, and the induction time in crystallization. Therefore, finding J is a central problem in recent nucleation theory.[6] Although it is very useful in all aspects, isotropic building blocks are mainly used in the three dimensional condition.

On the other hands, the effect of anisotropic interactions on the nucleation barrier and/or rate have received only a little attention.[7-10] According to the classical nucleation theory (CNT), the essential in the processes of crystal nucleation, growth, and dissolution can be unveiled even by the simplest model.[11] However, nucleation theory with anisotropic building block is insufficient to be explained via preceding models. The nucleation of two-dimensional(2D) on their own substrate is a simple prototype that allows investigating the effect of anisotropic intermolecular interactions on the nucleation behavior.[12,13] To date, 2D crystallization theory originated from exotic nucleation of the anisotropic building block (i.e. amyloid fiber) has proposed the concept of meta-nucleation and contributed to widening our understanding of nucleation.[14-16] Yet, a lot of nucleation researches about crystallization and nucleation with the anisotropic building block are usually based on heteronucleation. Recently, there is a recent paper that shows good insights about nucleation in homogeneous system with the anisotropic building block.[17] Although this explained the meta-nucleation of amyloid fibril, the thermodynamic interpretation of meta-nucleation theory is not described in a unified way of CNT.

We thereby demonstrate the validity of metanucleation from the experimental result. Isolation of nuclei of the Dirac metamaterial is experimentally conducted. These attempts have been made to isolate metanuclei and to experimentally identify its kinetics. When HYLION-12 is rapidly crystallized at 77K, an exotic cluster having only the z-axis periodicity of the Dirac metamaterial is obtained.[18] This is very similar phenomena of metanucleation of amyloid fibril. The unique nature of the ingredient of the Dirac metamaterial can help to explain the exotic thermodynamic aspect of the nucleation process with the anisotropic building block. Moreover, crystallization at various temperatures confirmed the nucleation and crystallization kinetics of HYLION-12 and it helps to the better understanding of the anisotropic effect on crystallization.

**Discovery of third phase of HYLION-12**

As the previous report suggests that HYLION-12 has two different solid phases; (i) the orthorhombic phase ($\Phi_o$): the first Dirac metamaterial and (ii) the Monoclinic phase ($\Phi_m$): a close-packed organic crystal.[18] Crystallization in various environments has been tried to form the Dirac metamaterial. Once the 1mM HYLION-12 solution in ethyl acetate is quenched in liquid nitrogen, the third phase of HYLION-12 is formed, which is different from the existing two phases and is designated as $\Phi_n$. The X-ray diffraction patterns of $\Phi_n$ show very high agreement with the patterns of $\Phi_o$, but the existence of a peak at 2.38° depicts differences between $\Phi_o$ and $\Phi_n$. (Figure 1a) It means that the (100) and (010) faces are only found in $\Phi_o$. Meanwhile, the peak at 19.58° from the (002) face, which indicates perpendicular-stacking of HYLION-12, can be found in both of $\Phi_o$ and $\Phi_n$. Therefore, $\Phi_n$ does not have a similar regularity of the Dirac metamaterial in the xy-plane and only has a characteristic unit cell for the z-axis. (Fig. 1b - 1d)

The XRD result shows a unique state of $\Phi_n$. It is impressive to have only the z-axis periodicity of the $\Phi_n$ without an amorphous region. The 1D crystal has not been found, and nanotube is only categorized as the quasi-1D crystal.[19] In addition, the term of 1D crystal is very contradictory itself because the definition of crystal is based on 'structural motifs' which may be atoms, molecules, or groups of atoms, molecules or ions in a 3D space lattice.[20,21] Therefore, it is very important to clarify the scientific meaning of $\Phi_n$.

The differential scanning calorimetry (DSC) analysis more clearly shows the relationship between $\Phi_o$ and $\Phi_n$. (Fig. 2a) The DSC curve of $\Phi_n$ represents two features that cannot be found in $\Phi_o$. One is the second-order phase transition at -23.15°C (glass transition temperature), large segmental motion with the movement of the long alkyl chains.[22] This temperature range is similar with polypropylene.[23] On the other hand, alkyl chains in $\Phi_o$ are entangled each other to form the xy-plane. Therefore, free motion of the alkyl chain is impossible, and $T_g$ was not observed in $\Phi_o$. The second interesting aspect is the first-order phase transition at 20.39°C. After this, the DSC curve starting from $\Phi_n$ shows the identical pattern of $\Phi_o$. Both of them exhibit exactly the same pattern and energy exchanging in the first cooling and second heating cycles (Figure 2b and 2c). In sum then, it can be inferred that $\Phi_n$ changed to $\Phi_o$ after the phase transition at 20.39°C.

Since the phase transition from $\Phi_n$ to $\Phi_o$ is irreversible and $\Phi_n$ has only the z-axis of the $\Phi_o$. $\Phi_n$ is regarded as a precursor of $\Phi_o$. According to the Ostwald's rule, the first phase in crystallization is not the most stable phase but one with the lowest energy barrier phase.[24] And it is reported that nuclei may undergo a first-order phase transition to become crystalline.[25] According to this, $\Phi_n$ can be regarded as nuclei of $\Phi_o$. Because $\Phi_n$ is neither a crystal with high crystallinity nor an amorphous solid.

To confirm the relationship between $\Phi_o$ and $\Phi_n$, grain sizes of crystals of HYLION-12

crystallized at various temperatures were analyzed. The XRD patterns of crystals at various temperatures depict the effect of temperature on the crystallization process of HYLION-12 (Figure 3a). The crystal of HYLION-12 is obtained under -63 °C and do not reveal the identified peak at 2.38°. As temperature increased, the peak at 2.38° gradually appears and a signature of $\Phi_o$ was observed in crystallization over -35 °C. Herein, $\Phi_n$ has a lower energy barrier than $\Phi_o$.[26-28] In other word, the energy gain from π-π interactions between HYLION-12 is larger than that from the van der Waals interactions. The Scherrer equation was used to calculate the grain size (Fig. S1 - S10 and Table. S1) and extrapolation was conducted.[29] (Fig. 3b) The positive intercept of the (002) face means existence of the z-axis stacking regardless of temperature. On the other hand, the (100) and (010) faces can be generated above -251°C. This is verified that $\Phi_n$ has lower energy barrier than $\Phi_o$. And the crystallite size on the z-axis increased smoothly, while that on the xy-plane increased steeply as the temperature increased. This suggests that the first Ostwald step in $\Phi_o$ started with z-axis stacking. Therefore, $\Phi_n$ is not another phase of HYLION-12 but rather is a preliminary step to form $\Phi_o$.

A low-temperature powder X-ray diffractometer (LT-XRD) was used to more precisely determine the relationship between $\Phi_n$ and $\Phi_o$. $\Phi_o$ can be formed after temperature over 12 °C and the (100) face gradually increases. (Fig. 4a) The relationship between $\Phi_n$ and $\Phi_o$ can be suggested by the change in crystallite size for (100) and (002) faces during the phase transition. The Scherrer equation was used to calculate the crystallite size before, (Fig S11-S15 and Table. S2) and after the phase transition. (Fig. 4d) Before the phase transition, $\Phi_n$ had only a (002) face, and the crystallite size was 52.8 nm and the crystallite size for the z-axis slightly decreased to 46.27 nm after the phase transition. On the other hand, the crystallite size for the xy-plane greatly increased to 54.32 nm. The increase of the xy-plane without a large loss of the z-axis indicates that $\Phi_n$ is the basic unit for forming $\Phi_o$. Furthermore, the observation of (100) face at 12 °C represents that $\Phi_o$ is spontaneously formed without going through a phase transition.

(Fig. 4a inset)

Crystallization consists of two stages: nucleation and growth.[30-32] The highlighted difference between nucleation and growth is the presence of reversibility and spontaneity. Nuclei are formed at the end of endlessly repeated assembly and disassembly of atoms or molecules and once a nucleus is formed, and reversibility is lost, growth spontaneously occurs in the saturated solution. The previous experiment also shows the direct relationship between $\Phi_n$ and $\Phi_o$. However, it is still difficult to distinguish whether it is a growth of nuclei or a solid to a solid phase transition. Thus, if $\Phi_n$ is the nuclei of $\Phi_o$, it spontaneously generates $\Phi_o$ from saturated solution in low temperature. To demonstrate this, $\Phi_n$ was contained in a saturated solution of HYLION-12 at -65°C, which is the only temperature that can be used to obtain $\Phi_n$ and the Scherrer equation was used to calculate the grain size. (Fig. S16-S19 and Table. S3) After two days, $\Phi_n$ naturally generates $\Phi_o$. Therefore, the assembly process of $\Phi_o$ from $\Phi_n$ is irreversible and spontaneous, and it would be regarded as growth. (Fig. 4b) Moreover, the change of crystallite size gives further confidence that it is growth. Although the crystallite size of the *z*-axis was maintained, the crystallite size of the xy-plane with the growth curve is increased from zero to 67nm for three days. (Figure 4e).

The cumulative effect of the double-stranded alkyl chain of HYLION-12 can be thought as the driving force of the growth and it describe the typical process of crystallization with the highly anisotropic building block.[33] When HYLION-12 does not sufficiently stack along the *z*-axis, it is energetically unfavorable to assemble in the xy-plane. Rather, the molecules accumulated to the z-axis provide the greater stabilization. However, if more than eight HYLIONs-12 are stacked on the *z*-axis, forming the xy-plane offers higher stability. It is because four stacked HYLION-12 particle (which have nine HYLIONs-12 on the *z*-axis) yield an energy benefit of approximately 19.04 kcal/mole when assembled in the xy-plane.[34] In different circumstances, stacking a HYLION-12 on the *z*-axis can yield an energy advantage of approximately 13.21 kcal/mole from π–π interactions.[35] Thus, the sum of energy gains from the alkyl chain surpasses the perpendicularly-stacked energy of the pyrene moiety, resulting in spontaneous conversion to $\Phi_o$.

It is somewhat difficult to admit that $\Phi_n$ represents a nuclei based on traditional nucleation theory.[36] Because the nuclei are assumed to be in an unstable transition phase resembling rolling stones that are ready to "fall over" to a stable crystal state. However, the previous experiment represents the possibility of $\Phi_n$ as a nucleus of $\Phi_o$. It means that the status of the nucleus changes from the transition to the intermediate state. And nonclassical crystallization has been recently studied in mesocrystal, suggesting the possibility of it.[37] It can occur with exotic building blocks which have a very strong anisotropic interaction, as in HYLION-12. Nonclassical crystallization with anisotropic interaction building blocks has been reported.[38,39] In particular, studies on amyloid fibrils have been progressing steadily to solve various diseases that are difficult to solve as well as have anisotropic interaction.[40,41] Although these researches increase the understanding the other aspect of crystallization, experimental implementation is intricated. However, isolation of $\Phi_n$ and various experiments demonstrate the possibility of non-classical crystallization. It can be possible for strong anisotropic interaction of HYLION-12 on the different axes. And crystallization of HYLION-12 can be used as a model molecule to experimentally implement the phenomenon.

On the other hands, the large scale fabrication method of metamaterial is a very important issue.[42] As mentioned above, $\Phi_n$ is regarded as the nuclei of $\Phi_o$ which is the first Dirac metamaterial. Therefore, large-scale metamaterials is fabricated from a pressure-induced method which is a well-known method of forming crystals from the nucleus.[43] Pressure-induced growth of $\Phi_o$ from $\Phi_n$ under ~ 8 GPa hydrostatic pressure rebalances the interparticle forces within $\Phi_n$ arrays and transforms them into a metamaterial ($\Phi_o$) (Fig. 4c) and the Scherrer equation was used to calculate the grain size. (Fig. S20 - S22 and Table. S4) At 3.84 GPa, a new peak appeared at 2.3°, indicating the creation of $\Phi_o$. As the pressure increased to 7.69 GPa, the intensity of the peak increased at 2.3 °. In this procedure, the crystallite size of the z-axis was maintained, and that of the xy-plane greatly was increased (Fig. 4f).

**Modification of classical nucleation theory (MCNT)**

In CNT, only the total numbers of bulk and surface interactions are considered. For example, in the colloidal model, the equation $\Delta G = \Delta\mu k^3 + \Gamma k^2$ is a widely accepted expression of nucleation.[44] This equation is based on the concept that the bulk and surface energies contribute to stabilization and destabilization, respectively. Following that, the differences in Gibbs free energy between $\Phi_n$ and $\Phi_o$ can be calculated by evaluating the dangling bonds (for destabilization) and connected bonds (for stabilization). When the $x$ numbers of HYLIONs-12 are stacked along the $z$-axis, $(x-1)$ π–π interactions can be formed for the bulk energy, and the top and bottom of the particles make surface energy. Bulk and surface energy can also be derived from the alkyl chains. When the $n^2$ numbers of perpendicularly-stacked particles that have the $x$ numbers of HYLIONs-12 stacked to the $z$-axis generate $\Phi_o$, $(n^2-n)(2x-1)$ bulk energy is produced from the overlapping alkyl chains. In this case, $2n(2x-1)$ surface energy is associated with the dangling edge of the alkyl chains. The sum of the bulk energy is represented by Eq. (1), and the sum of the surface energy is given by Eq. (2):

$$E_{bulk}(x,n) = n^2(x-1)E_{\pi-\pi} + (n^2-n)(2x-1)E_{alkyl\ chain} \qquad (1)$$

$$E_{surface}(x,n) = 2nE_{\pi-\pi} + 2n(2x-1)E_{alkyl\ chain} \qquad (2)$$

The π–π interaction energy of pyrene is 13.21 kcal/mole and gain energy of overlapping alkyl chains is about 0.28 kcal/mole. Four alkyl chain interactions exist in a unit cell; x and n are converted to the scale of length, yielding a single-variant equation. The internal molecular distance is 0.45 Å, and the cell parameters of the x and y-axes are 34 Å and 37 Å, respectively. Therefore, Eqs. (1) and (2) can be converted to an equation that has length as its single variable.

$$E_{bulk}(l) = 1.95l^3 - 13.93l^2 + 0.56l \quad (3)$$

$$E_{suface}(l) = 0.54l^2 + 25.28l \quad (4)$$

The total energy of the crystal during nucleation consists of the bulk energy required to stabilize the structure. The surface energy is needed to destabilize the structure. Defining, the total change in Gibbs energy of HYLION-12 related to nucleation is as follows:

$$\Delta G(l) = -1.949l^3 + 14.4697l^2 + 24.72l \quad (5)$$

The graph of Eq. (5) well represents classical nucleation theory, where there is an energy barrier (Figure 5a). According to traditional nucleation theory, $\Phi_n$ cannot be isolated. In the absence of anisotropic interaction effect, Eq. (5) cannot explain the isolation of $\Phi_n$. The existing nucleation models are based on spherical particles, where there is no energy difference based on the direction. One of the biggest differences in nucleation of molecules from colloids is the energy differences in various directions.

In classical nucleation theory, Gibbs free energy has two components; bulk and surface energies. The bulk energy is proportional to the volume of the nucleus, which is assumed to be spherical and so is related to the cube of the radius. The surface term is proportional to the surface area of the nucleus and so is related to the square of the radius. As mentioned above, nuclei in real molecules have different interaction energies in different directions, resulting in a preferred orientation. That is why volume and surface alone are not sufficient to reflect the actual nucleation of molecules. Thus, the third term is related only to the radius and represents the anisotropic interaction ratio as an anisotropic factor, in the molecule,[33] represented as $\Delta G =$

$\Delta \mu k^3 + \Gamma k^2 - \xi k.$

$$\Delta G(l) = -1.95l^3 + 14.47l^2 + (24.72 - \xi)l \quad (6)$$

To show the effect of $\xi$, Equation (6) was plotted for various values of $\xi$ (Figure 5a). Introduction of an anisotropic term has a major impact on the relationship between nucleation and growth. The Gibbs free energy curve of nucleation and growth without the anisotropic term clearly shows one transition state. However, the height of the Gibbs free energy curve decreases gradually, and the starting direction moves downward as $\xi$ increase. In Equation (6), the critical value of the anisotropic constant is 24.72. For values smaller than that, it shows a positive slope at the initial state and gradually increases to the transition state. For values larger than that, it shows a negative slope and should be an intermediate state preceding the transition state. Moreover, if the value of $\xi$ is fairly large, it was found to be an intermediate state and no longer a transition state.

For HYLION-12, the perpendicularly-stacked energy that comes from the enlarged quadrupole effect is approximately 13.21 kcal/mole[35], and the overlapping energy of alkyl chains is approximately 0.28 kcal/mole[34], so the value of $\xi$ is 47.19. If that of $\xi$ is substituted at Eq. (6), the new Gibbs energy equation that reflects the anisotropic interaction energy, can be rewritten to the following Eq. (7).

$$\Delta G(l) = -1.94l^3 + 14.47l^2 - 22.47l \quad (7)$$

In Eq. (7), the valley occurs before the energy barrier (see Figure 5b). The critical nucleation energy barrier graph created from Eq. (7) suggests that stable nuclei are formed when the nucleating molecule has enough anisotropic interaction energy, $\xi$. This graph also shows the process of spontaneous crystal growth from the nucleated molecules of HYLION-12 stacked along the z axis to the orthorhombic phase grown to the xy - plane direction.

The maximum Gibbs energy barrier in Eq. (7) is 16.9 kcal/mole, but the interaction energy from the alkyl chain overlap can overcome this barrier spontaneously when nine HYLIONs-12 are stacked along the $z$-axis and four stacked particles generate $\Phi_o$. It means that $\Phi_n$ can overcome the energy barrier through accumulation of van der Waals interactions. So, $\Phi_n$ is cautiously proposed as a nucleation state of $\Phi_o$.

**Conclusion**

$\Phi_n$ discovered newly is a very unique phase with only z-axis characteristics of the orthorhombic phase of a Dirac metamaterial. It is difficult to classify as a crystal because it does not have three-dimensional regularity and periodicity and shows repeatability only in an axial direction. However, it has been clearly isolated and fully characterized using DSC and LT-XRD, and the results suggest that it may indicate the nuclei of the Dirac metamaterial. In addition, three distinctive experiments involving crystallization at various temperatures, growth of the orthorhombic phase from the third phase in a saturated solution of HYLION-12, and pressure-induced crystallization further supported it as nuclei of the Dirac metamaterial. Furthermore, a new crystallization equation involving the anisotropic effect of HYLION-12 is proposed to explain isolation of crystals with the third phase that has not been reported so far by the existing crystallization theory. The finding is also significant in terms of providing a new strategy for forming a large-area metamaterial through the growth of isolated nuclei. The work also demonstrates the validity of the new crystallization theory including anisotropic effects. This approach is expected to contribute to the development of the related field by providing a new method which is efficient for obtaining better crystals on a large area

**References**

1. Kashchiev, D. *Nucleation*. (Elsevier, 2000).
2. Vekilov, P. G. Nucleation. *Crystal growth & design* **10**, 5007-5019 (2010).
3. Wolde, P.-R. Simulation of homogeneous crystal nucleation close to coexistence. *Faraday discussions* **104**, 93-110 (1996).
4. Turnbull, D. & Fisher, J. C. Rate of nucleation in condensed systems. *The Journal of chemical physics* **17**, 71-73 (1949).
5. Auer, S. & Frenkel, D. Prediction of absolute crystal-nucleation rate in hard-sphere colloids. *Nature* **409**, 1020 (2001).
6. Kashchiev, D. On the relation between nucleation work, nucleus size, and nucleation rate. *The Journal of chemical physics* **76**, 5098-5102 (1982).
7. Kaischew, R. Equilibrium shape and work of formation of crystalline nuclei on a foreign substrate. *Commun. Bulg. Acad. Sci* **1**, 100 (1950).
8. Auer, S., Dobson, C. M., Vendruscolo, M. & Maritan, A. Self-Templated Nucleation in Peptide and Protein Aggregation. *Physical Review Letters* **101**, 258101, doi:10.1103/PhysRevLett.101.258101 (2008).
9. Quigley, D. & Rodger, P. M. Metadynamics simulations of ice nucleation and growth. *The Journal of chemical physics* **128**, 154518, doi:10.1063/1.2888999 (2008).
10. Duff, N. & Peters, B. Nucleation in a Potts lattice gas model of crystallization from solution. *The Journal of chemical physics* **131**, 184101, doi:10.1063/1.3250934 (2009).
11. Stranski, I. & Kaischew, R. Crystal growth and crystal nucleation. *Z. Phys* **36** (1935).
12. Kossel, W. L. J. *Die molekularen Vorgänge beim kristallwachstum*. (1928).
13. Stranski, I. N. Zur theorie des kristallwachstums. *Zeitschrift für physikalische Chemie* **136**, 259-278 (1928).
14. Cabriolu, R., Kashchiev, D. & Auer, S. Atomistic theory of amyloid fibril nucleation. *The Journal of chemical physics* **133**, 225101, doi:10.1063/1.3512642 (2010).
15. Arosio, P., Knowles, T. P. & Linse, S. On the lag phase in amyloid fibril formation. *Physical Chemistry Chemical Physics* **17**, 7606-7618 (2015).
16. Trusova, V. & Gorbenko, G. MODELIZATION OF AMYLOID FIBRIL SELF-ASSEMBLY. *East European Journal of Physics* **5**, 47-54 (2018).
17. Kashchiev, D., Cabriolu, R. & Auer, S. Confounding the Paradigm: Peculiarities of Amyloid Fibril Nucleation. *Journal of the American Chemical Society* **135**, 1531-1539, doi:10.1021/ja311228d (2013).
18. Choi, K. H., Hwang, D. Y. & Suh, D. H. Dirac Metamaterial Assembled by Pyrene Derivative and its Topological Photonics. *arXiv preprint arXiv:2101.04359* (2021).
19. Novoselov, K. *et al.* Two-dimensional atomic crystals. *Proceedings of the National Academy of Sciences* **102**, 10451-10453 (2005).
20. Kittel, C., McEuen, P. & McEuen, P. *Introduction to solid state physics*. Vol. 8 (Wiley New York, 1996).

21   Atkins, P. W. & De Paula, J. *Physical chemistry*. Vol. 4 (Oxford university press, Oxford UK, 1998).

22   Bershteĭn, V. A. & Egorov, V. M. *Differential scanning calorimetry of polymers: physics, chemistry, analysis, technology*.   (Prentice Hall, 1994).

23   Charles, E., Charles, A., James, W. & Mark, T.    (Hanser Publications: Cincinnati, OH, USA, 2005).

24   Ostwald, W. File: Wilhelm Ostwald-Studien über die Bildung und Umwandlung fester Körper. pdf-Fuhz Articles. *Zeitschrift für Physikalische Chemie* **22**, 289-330 (1897).

25   Oxtoby, D. W. Nucleation of first-order phase transitions. *Accounts of Chemical Research* **31**, 91-97 (1998).

26   Brailsford, A. & Wynblatt, P. The dependence of Ostwald ripening kinetics on particle volume fraction. *Acta Metallurgica* **27**, 489-497 (1979).

27   Voorhees, P. W. The theory of Ostwald ripening. *Journal of Statistical Physics* **38**, 231-252 (1985).

28   Ratke, L. & Voorhees, P. W. *Growth and coarsening: Ostwald ripening in material processing*.   (Springer Science & Business Media, 2013).

29   Drits, V., Srodon, J. & Eberl, D. XRD measurement of mean crystallite thickness of illite and illite/smectite: Reappraisal of the Kubler index and the Scherrer equation. *Clays and clay minerals* **45**, 461-475 (1997).

30   West, A. R. *Solid state chemistry and its applications*.   (John Wiley & Sons, 2013).

31   Elwell, D. Crystal growth from high-temperature solutions. (Scheel, Hans J, 2011).

32   Cravillon, J. *et al.* Controlling zeolitic imidazolate framework nano-and microcrystal formation: Insight into crystal growth by time-resolved in situ static light scattering. *Chemistry of Materials* **23**, 2130-2141 (2011).

33   Cabriolu, R., Kashchiev, D. & Auer, S. Breakdown of nucleation theory for crystals with strongly anisotropic interactions between molecules. *The Journal of chemical physics* **137**, 204903 (2012).

34   Shapiro, E. & Ohki, S. The interaction energy between hydrocarbon chains. *Journal of Colloid and Interface Science* **47**, 38-49 (1974).

35   Li, J. *et al.* Describing curved-planar pi-pi interactions: modeled by corannulene, pyrene and coronene. *Physical chemistry chemical physics : PCCP* **15**, 12694-12701, doi:10.1039/c3cp51095f (2013).

36   Mullin, J. W. *Crystallization*.   (Butterworth-Heinemann, 2001).

37   Cölfen, H. & Antonietti, M. Mesocrystals: inorganic superstructures made by highly parallel crystallization and controlled alignment. *Angewandte Chemie International Edition* **44**, 5576-5591 (2005).

38   Glotzer, S. C. & Solomon, M. J. Anisotropy of building blocks and their assembly into complex structures. *Nature materials* **6**, 557 (2007).

39   Jones, M. R. *et al.* DNA-nanoparticle superlattices formed from anisotropic building blocks. *Nature materials* **9**, 913 (2010).

40   Knowles, T. P. & Mezzenga, R. Amyloid fibrils as building blocks for natural and artificial functional materials. *Advanced Materials* **28**, 6546-6561 (2016).

41   Auer, S. Amyloid fibril nucleation: effect of amino acid hydrophobicity. *The Journal of Physical Chemistry B* **118**, 5289-5299 (2014).

42   Gwinner, M. C. *et al.* Periodic Large-Area Metallic Split-Ring Resonator Metamaterial Fabrication Based on Shadow Nanosphere Lithography. *Small* **5**, 400-406 (2009).

43   Schroer, M. A. *et al.* Pressure-Stimulated Supercrystal Formation in Nanoparticle Suspensions. *The journal of physical chemistry letters* **9**, 4720-4724 (2018).

44   Abraham, F. F. *Homogeneous nucleation theory*. Vol. 263 (Elsevier, 1974).

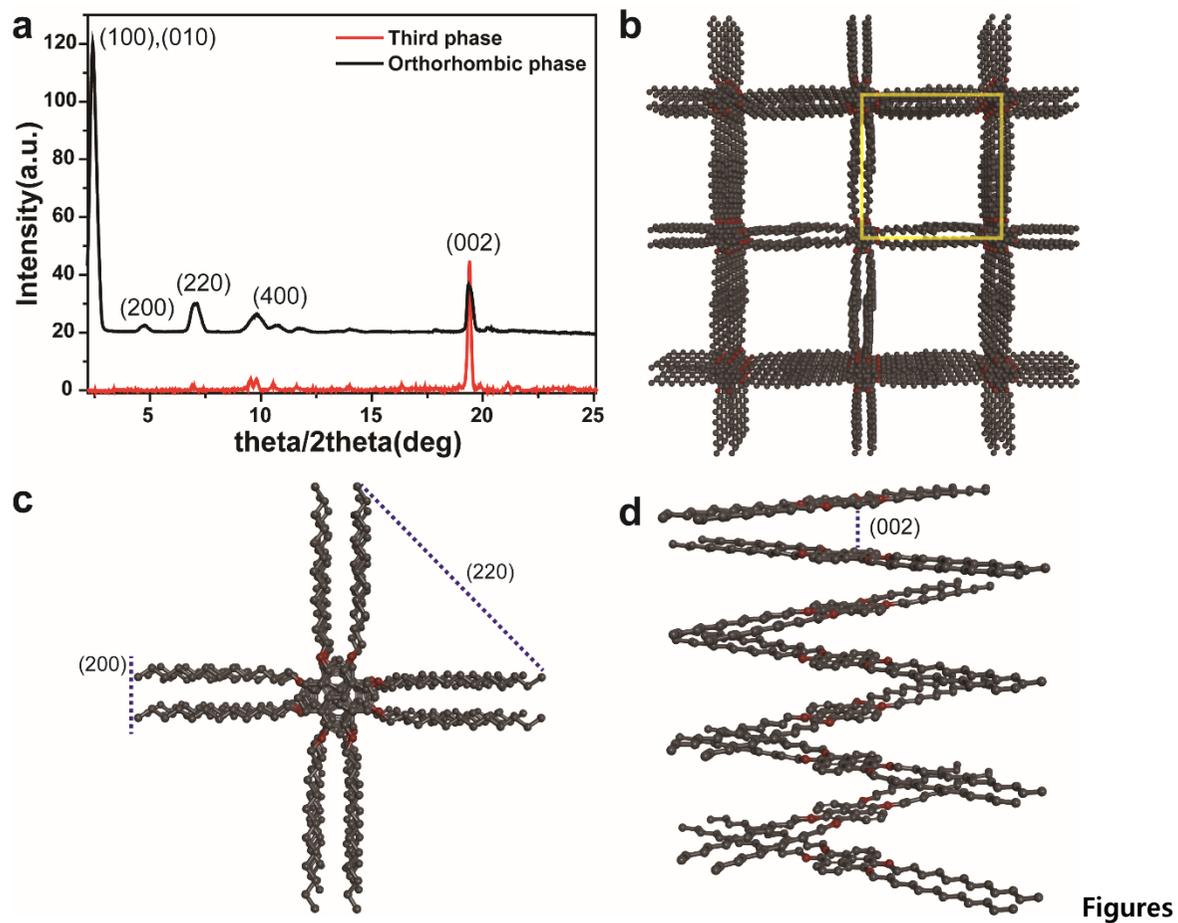

**Figure 1. Structural characterization of $\Phi_n$. a.** Powder x-ray diffraction pattern of $\Phi_n$. This clearly shows $\Phi_n$ only has the (002) face of $\Phi_o$. **b.** Molecular simulation of unit cell of $\Phi_o$. **c and d.** Molecular modeling of $\Phi_n$. It only has regularity and periodicity only for z-axis.

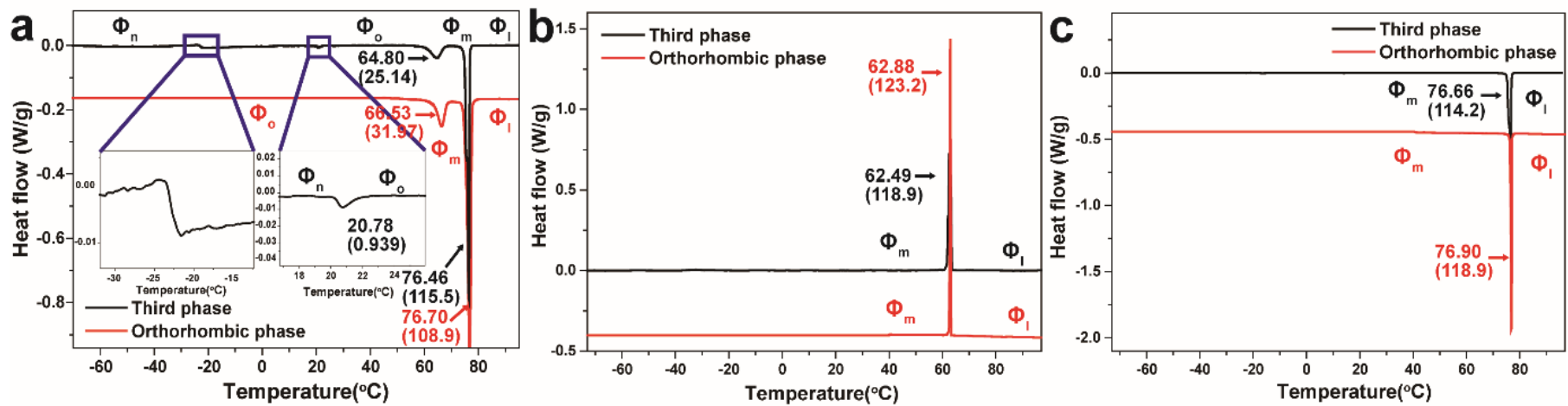

**Figure 2. Temperature-dependent phase transition from $\Phi_n$ to $\Phi_o$. a.** First heating cycle. This clearly depicts exotic characteristics of $\Phi_n$. Especially, second order phase transition of $\Phi_n$ at -23.23 °C shows unlinked alkyl chain. After first order phase transition at 20.78 °C, $\Phi_n$ have same DSC pattern of $\Phi_o$. **b.** First cooling cycle. **c.** Second heating cycle.

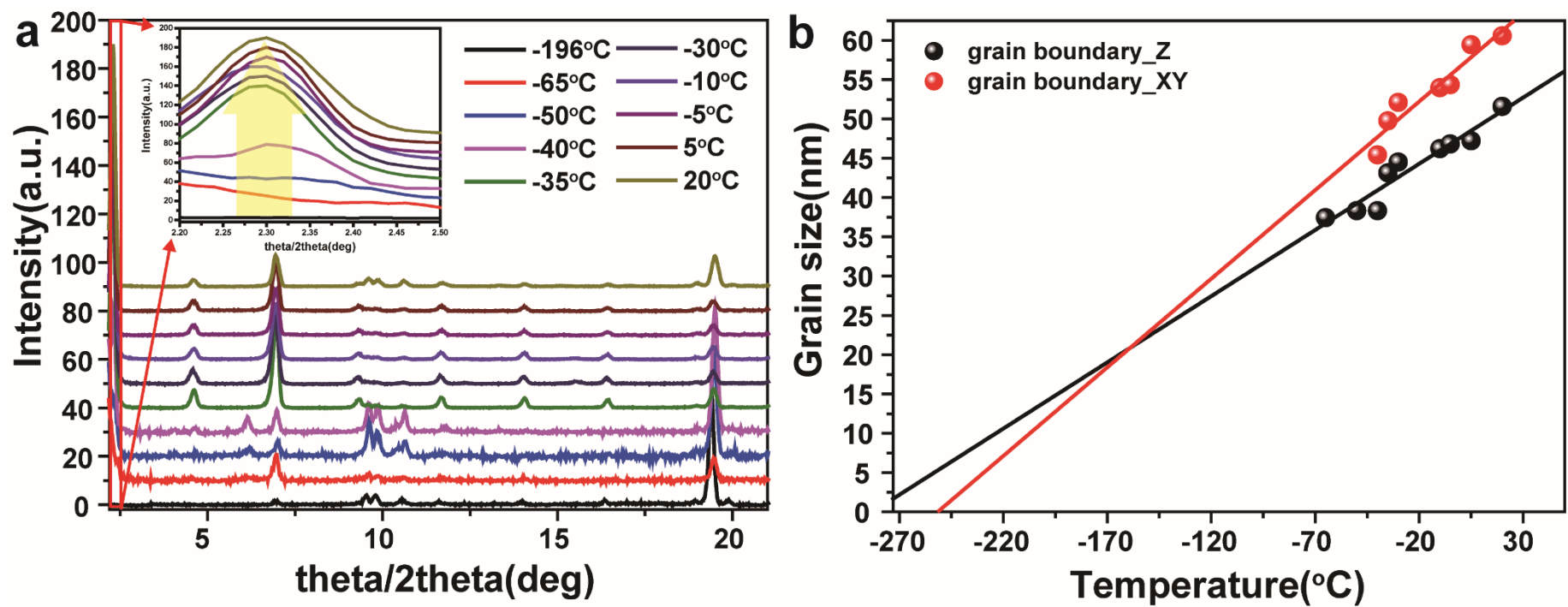

**Figure 3. Crystallization of HYLION-12 at various temperatures. a.** Powder x-ray diffraction patterns of various crystals of HYLION-12 crystallized at various temperatures **b.** Trend of crystallite size of HYLION-12 depends on crystallization temperature.

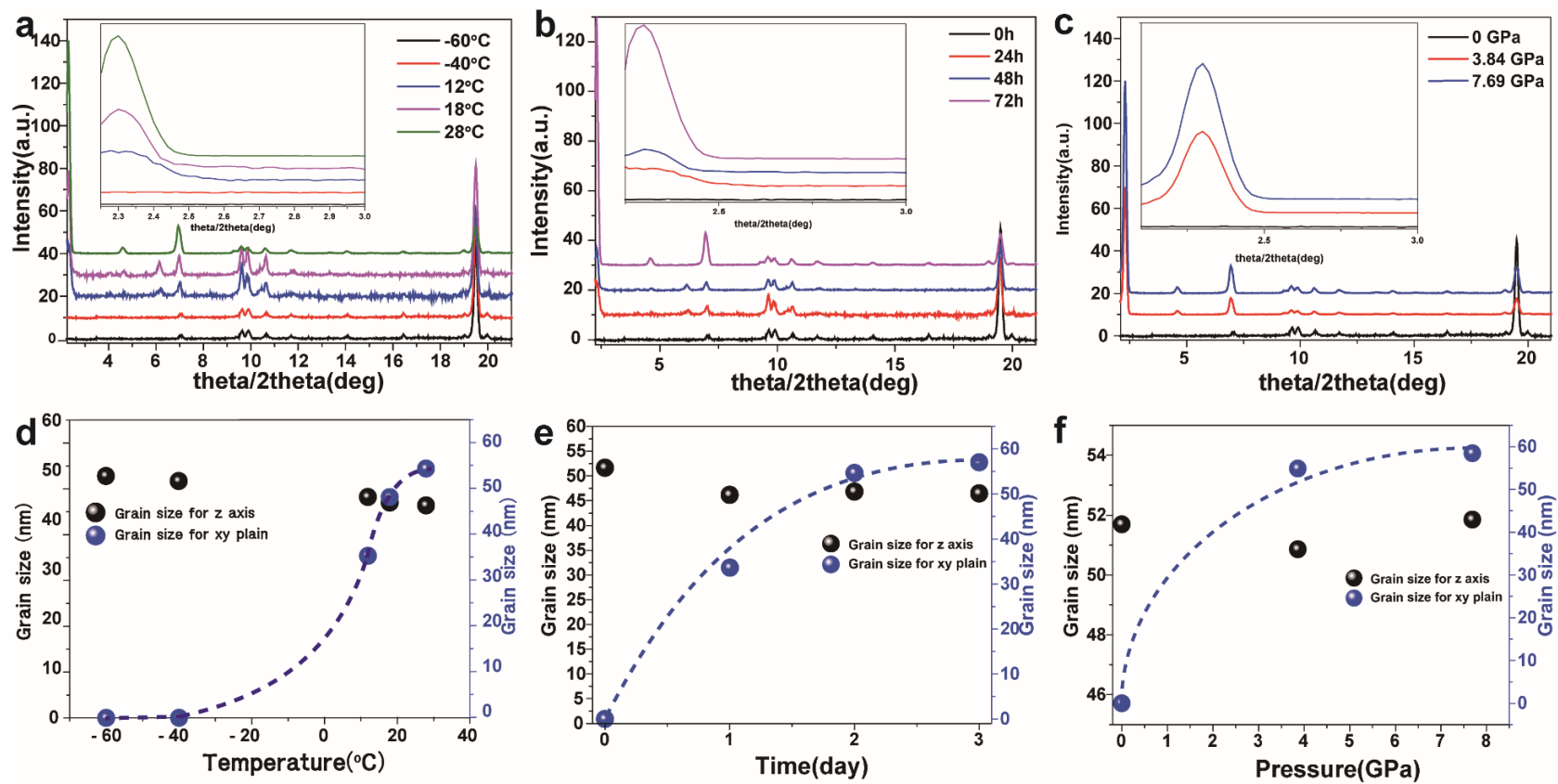

**Figure 4. Growth of $\Phi_n$ vs. conditions of temperature, time and pressure. a and d.** The PXRD pattern of $\Phi_n$ at different temperatures and grain size for each axis. This clearly shows that spontaneous growth for $\Phi_o$ does not occur at temperatures below glass transition temperature, but spontaneously becomes $\Phi_o$ even though it is lower than phase transition temperature. **b and e.** The PXRD pattern of $\Phi_n$ and its grain boundary size stored at 0 °C for one to three days. It also clearly represents growth in xy-plane without loss in z-axis. It demonstrates spontaneous generation of $\Phi_o$ from $\Phi_n$ like growth stage of crystallization. **c and f.** PXRD patterns of press-induced crystallization of $\Phi_n$ to $\Phi_o$ and its grain boundary size.

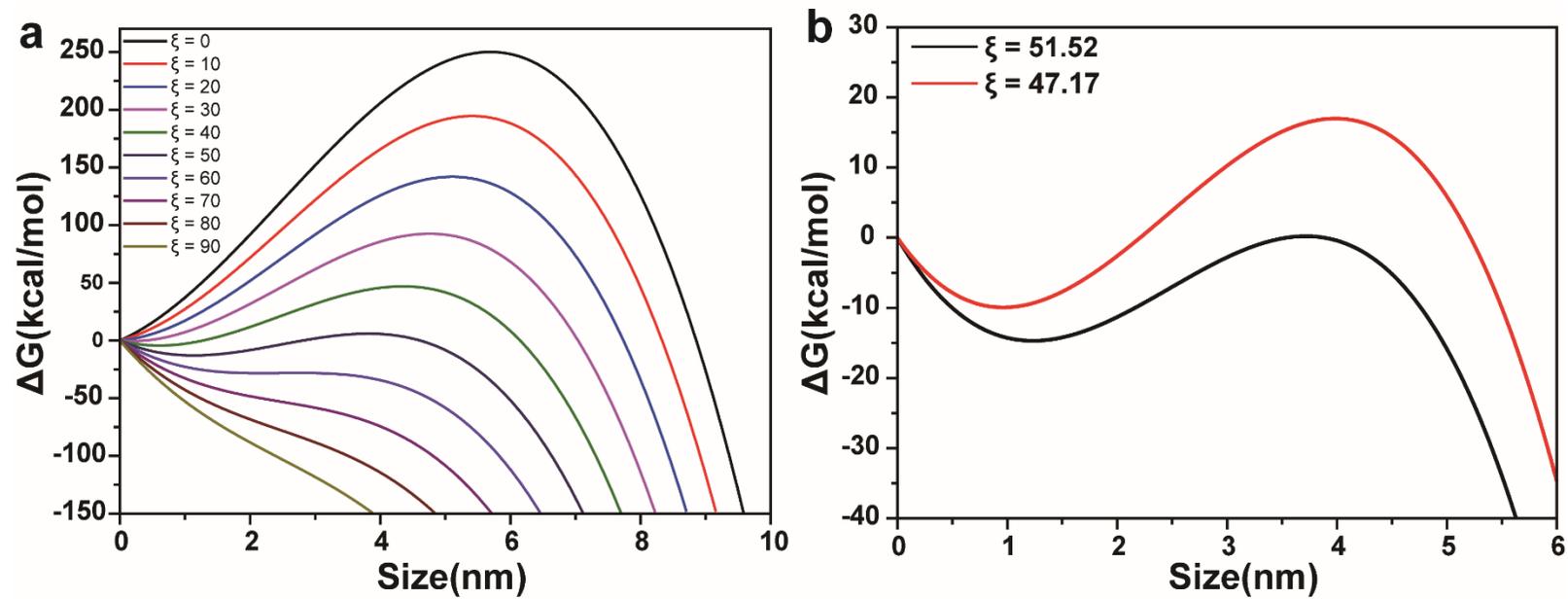

**Figure 5. Nucleation energy barriers. a.** Gibbs free energy barriers for various anisotropic factors **b.** Critical anisotropic factor and calculated anisotropic factor of HYLION-12.

**Experimental section**

**Materials**: Pyrene, NaIO$_4$, Bu$_4$NBr, Na$_2$S$_2$O$_4$, and dodecyl bromide were purchased from Sigma Aldrich and used as received. CDCl$_3$ was purchased from Cambridge Isotope Laboratories and was used for the $^1$H NMR spectroscopic studies according to a method reported in previous work.

**Crystallization**: the creation of the third phase of HYLION-12 was carried out through rapid quenching. A 100 μM HYLION-12 solution was prepared in ethyl acetate and was dipped into liquid nitrogen or an isopropyl alcohol bath at various temperatures. It was then filtered and freeze-dried after slowly dissolving at 0°C.

**Thermal analysis**: Thermal transitions were measured on a TA Instrument SDT Q 600 ver. 20.9 Build 20 system differential scanning calorimeter (DSC) integrated with a refrigerated cooling system (RCS). In all cases, the heating and cooling rates were 1 °C min$^{-1}$. The transition temperatures were measured based on the maxima and minima of their endothermic and exothermic peaks.

**X-ray diffractometer**: Theta/2theta were measured on a D/Max-2500 (Rigaku Co., Ltd.) diffractometer using monochromatized Cu-Kα (λ = 0.15418 nm) radiation operating at 40 kV and 100 mA. Low-temperature XRD was measured on a Bruker D8 Advance with a high-temperature attachment in the theta/2theta geometry. The low-temperature stage allowed the sample to be measured at tightly controlled temperatures from -60°C to 28°C.

**Scherrer equation:** The Scherrer equation was obtained by

$$\beta(\theta) = \frac{K\lambda}{B\cos\theta}$$

where K is a constant near unity, B is the full width at half maximum of the peak at theta, and λ is the wavelength of the X-ray, in this case 1.54178 Å.

**Supplementary information for:**

**Growth of metamaterial from Isolated nuclei with anisotropic building block**

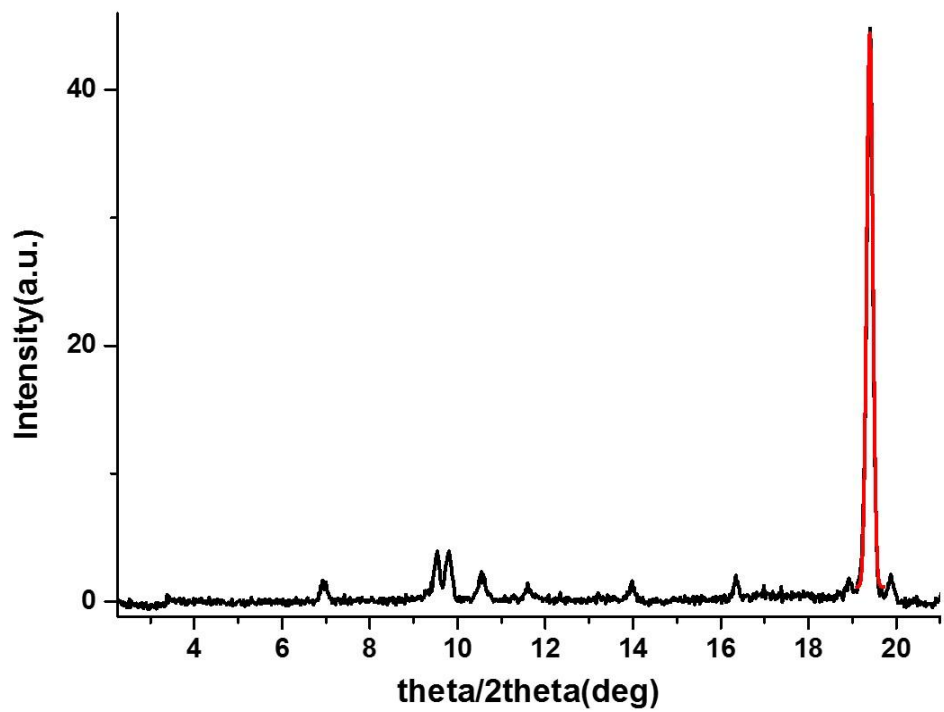

Figure S1. XRD pattern of HYLION-12 crysatllized at -196°C and fitting restuls for scherrer equation.

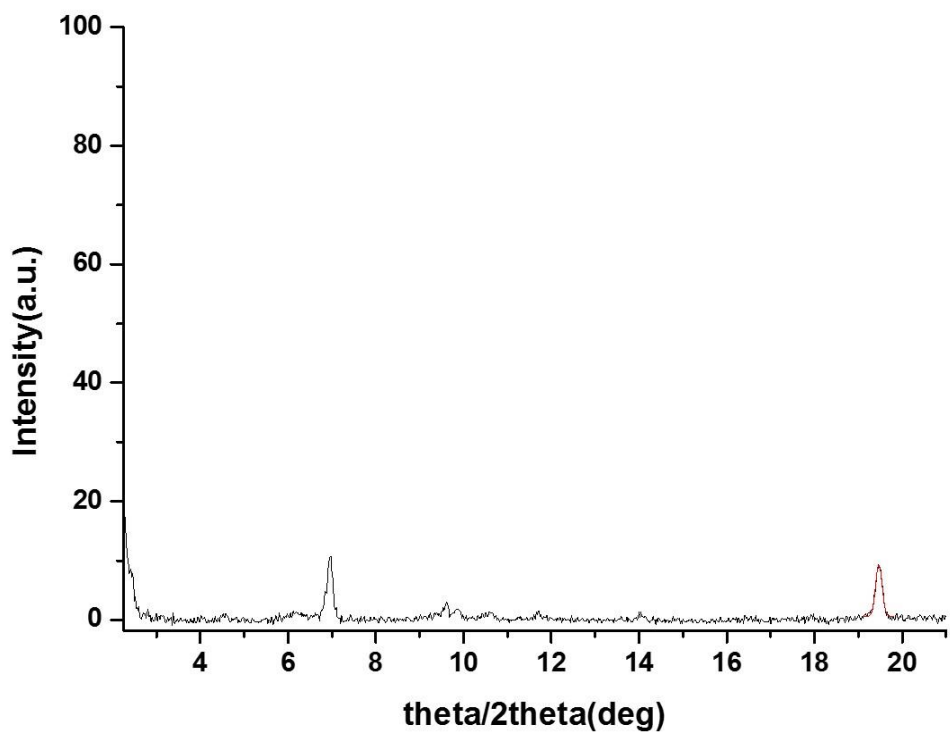

**Figure S2.** XRD pattern of HYLION-12 crysatllized at -65°C and fitting restuls for scherrer equation.

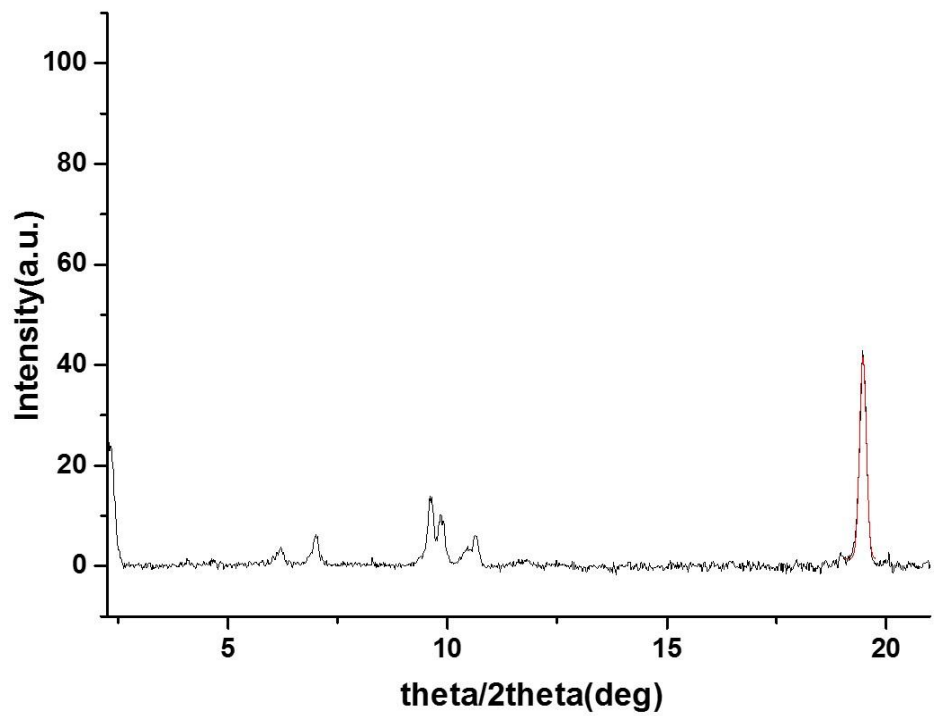

Figure S3. XRD pattern of HYLION-12 crysatllized at -50°C and fitting restuls for scherrer equation.

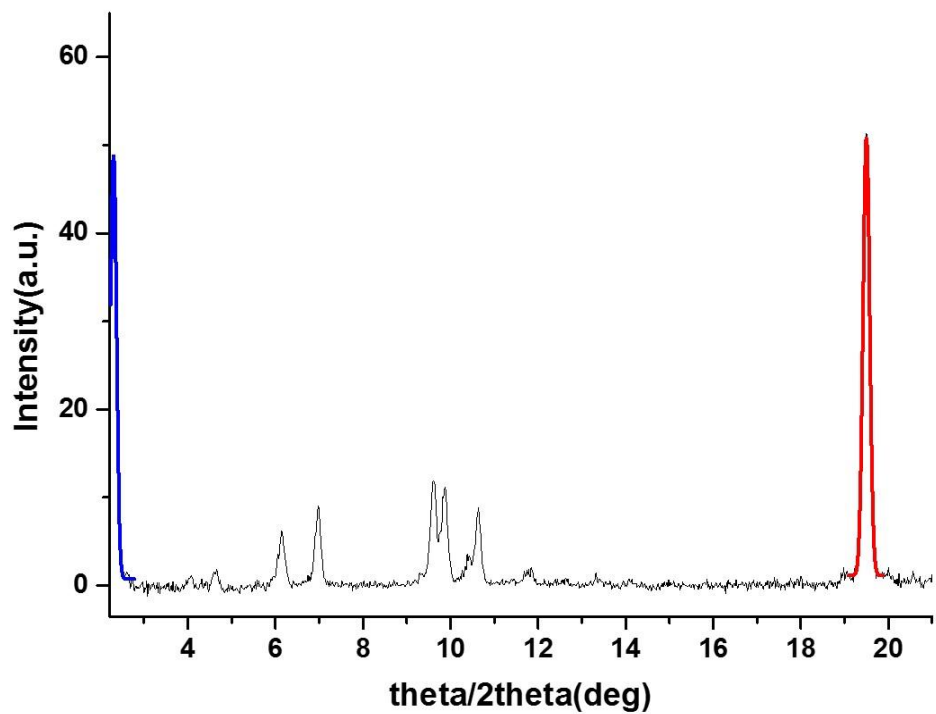

Figure S4. XRD pattern of HYLION-12 crysatllized at -40°C and fitting restuls for scherrer equation.

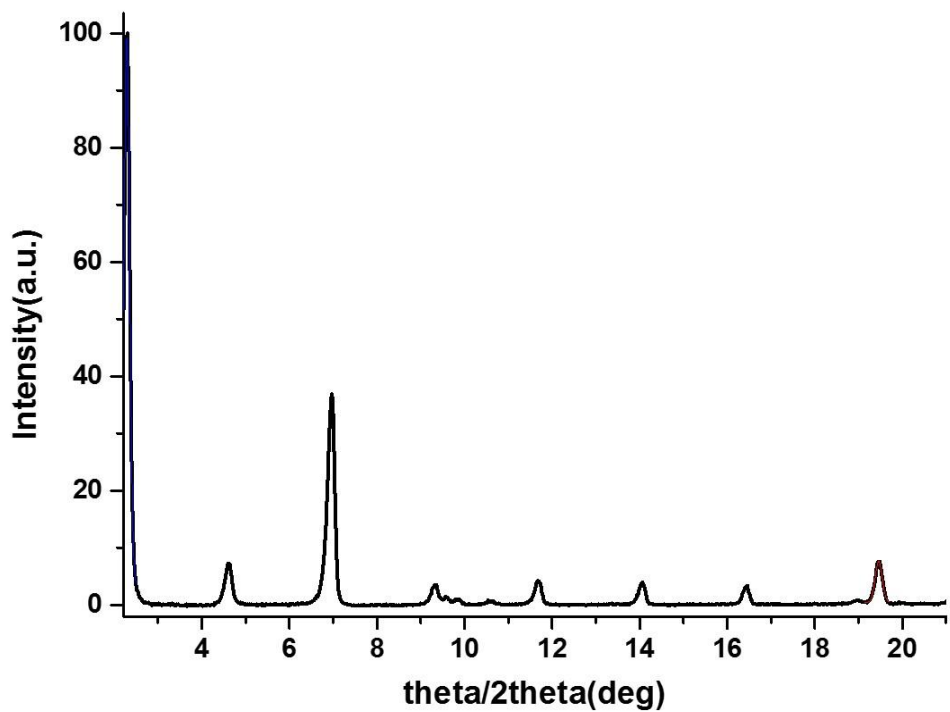

Figure S5. XRD pattern of HYLION-12 crysatllized at -35°C and fitting restuls for scherrer equation.

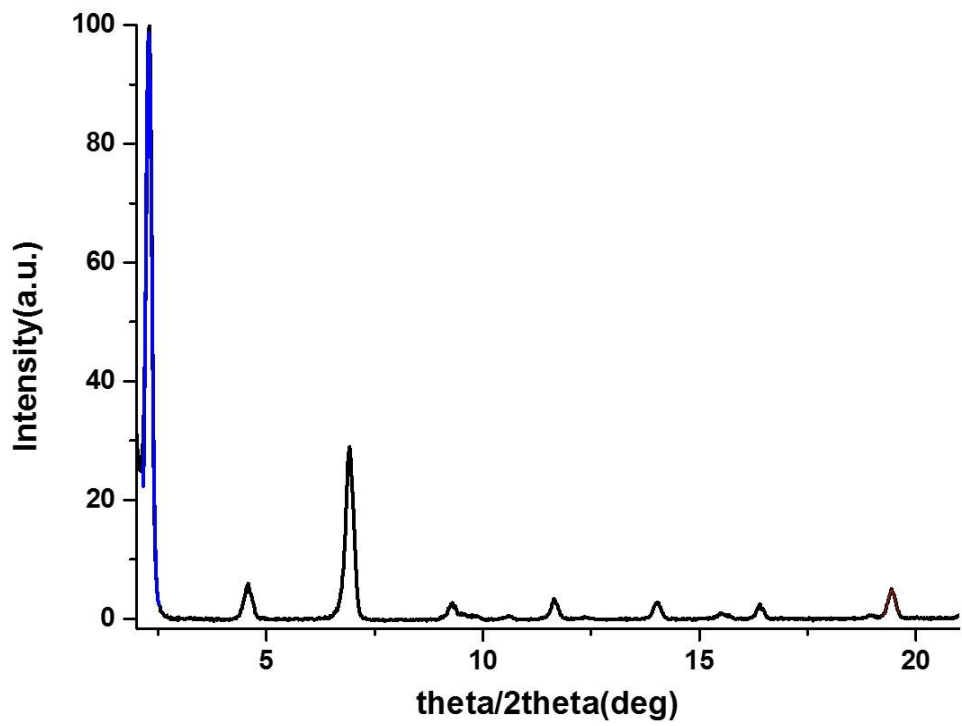

Figure S6. XRD pattern of HYLION-12 crysatllized at -30°C and fitting restuls for scherrer equation.

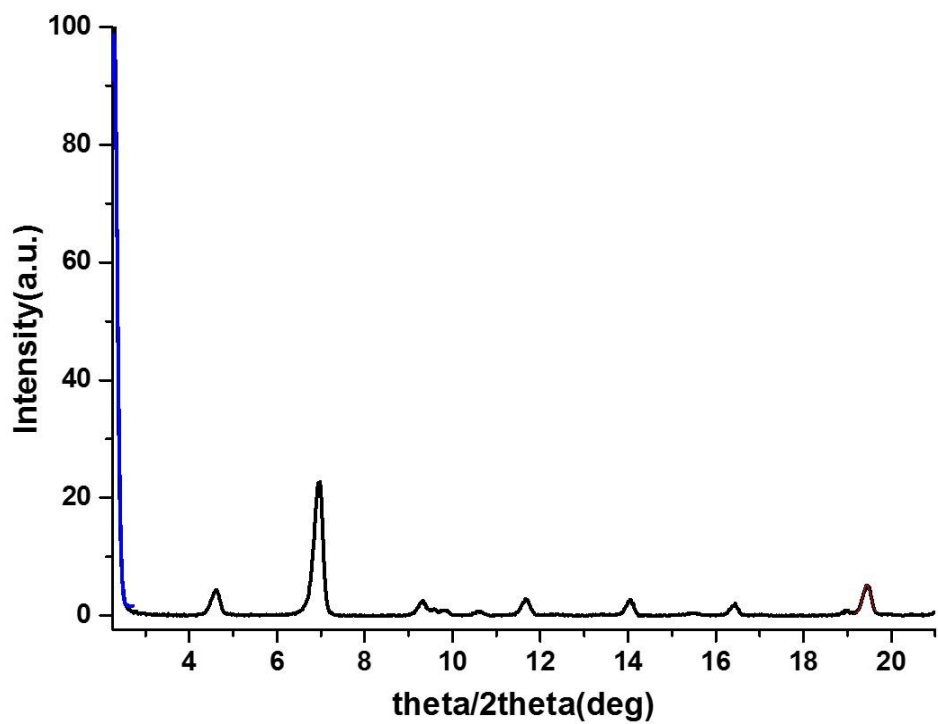

Figure S7. XRD pattern of HYLION-12 crysatllized at -10°C and fitting restuls for scherrer equation.

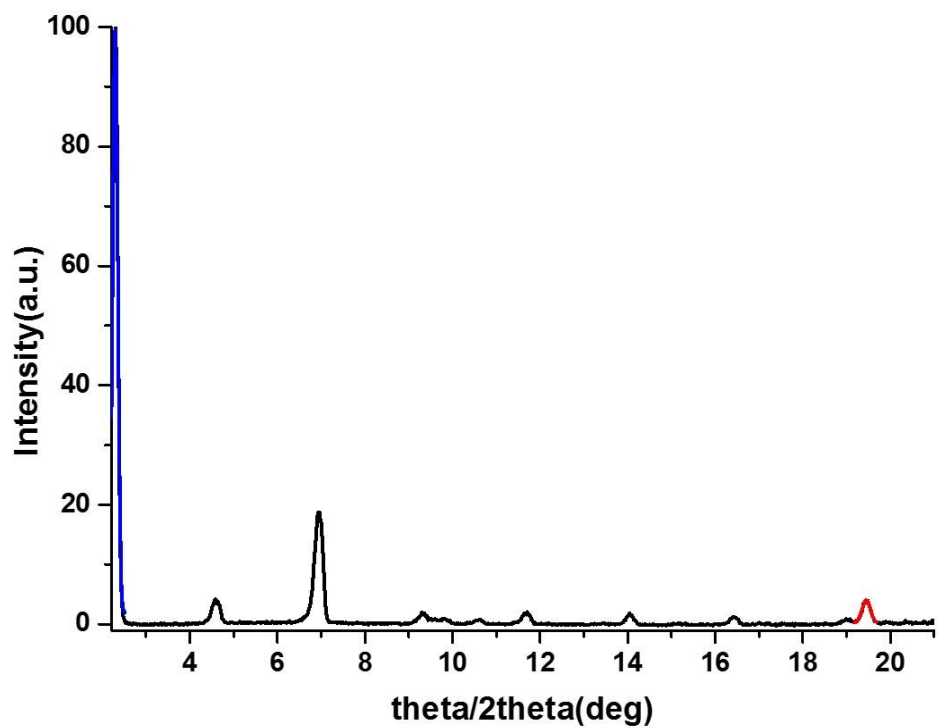

Figure S8. XRD pattern of HYLION-12 crysatllized at -5°C and fitting restuls for scherrer equation.

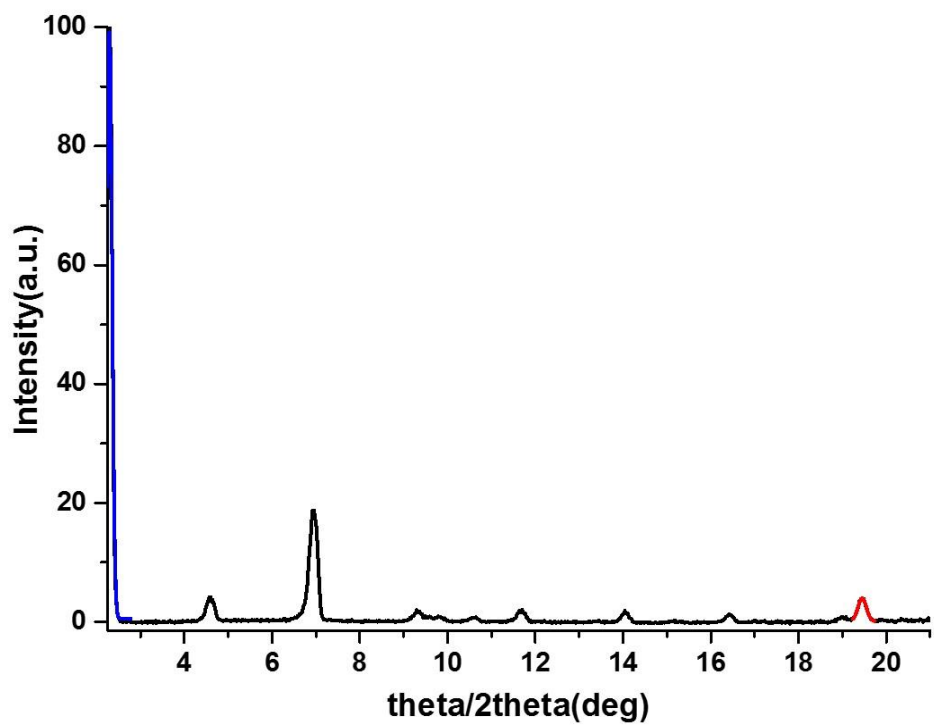

**Figure S9.** XRD pattern of HYLION-12 crysatllized at 5°C and fitting restuls for scherrer equation.

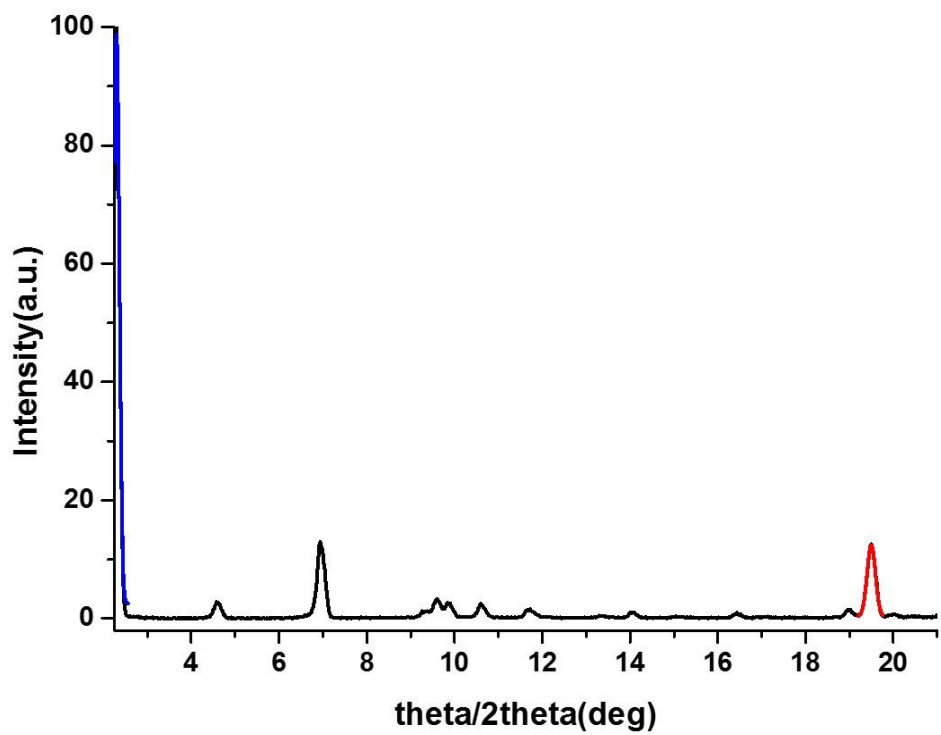

**Figure S10.** XRD pattern of HYLION-12 crysatllized at 20°C and fitting restuls for scherrer equation.

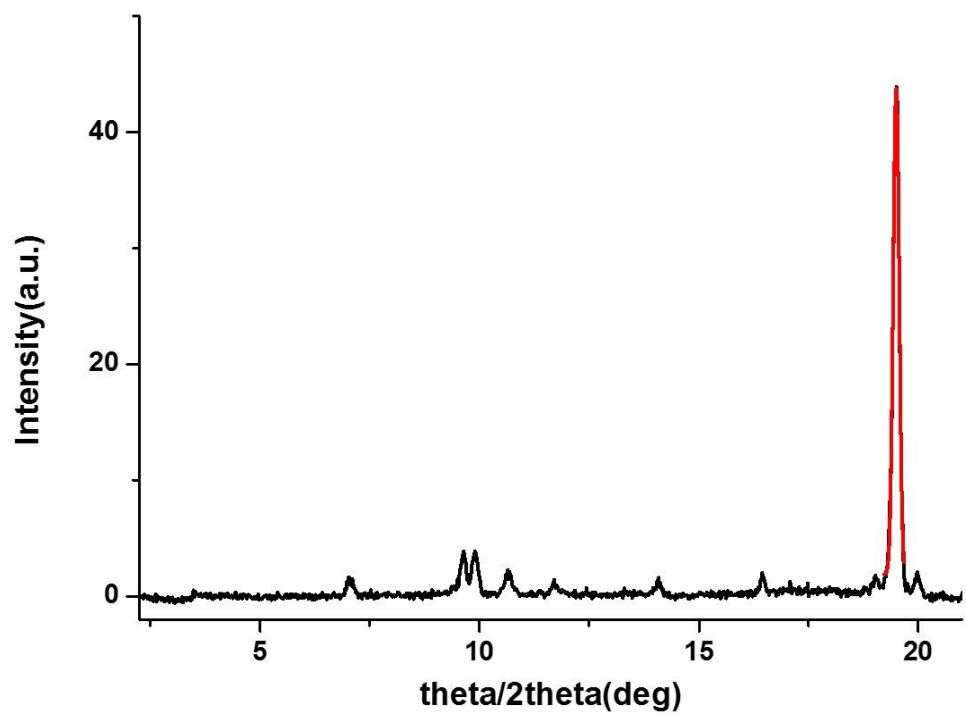

Figure S11. LT-XRD pattern measured at -60°C and fitting restuls for scherrer equation.

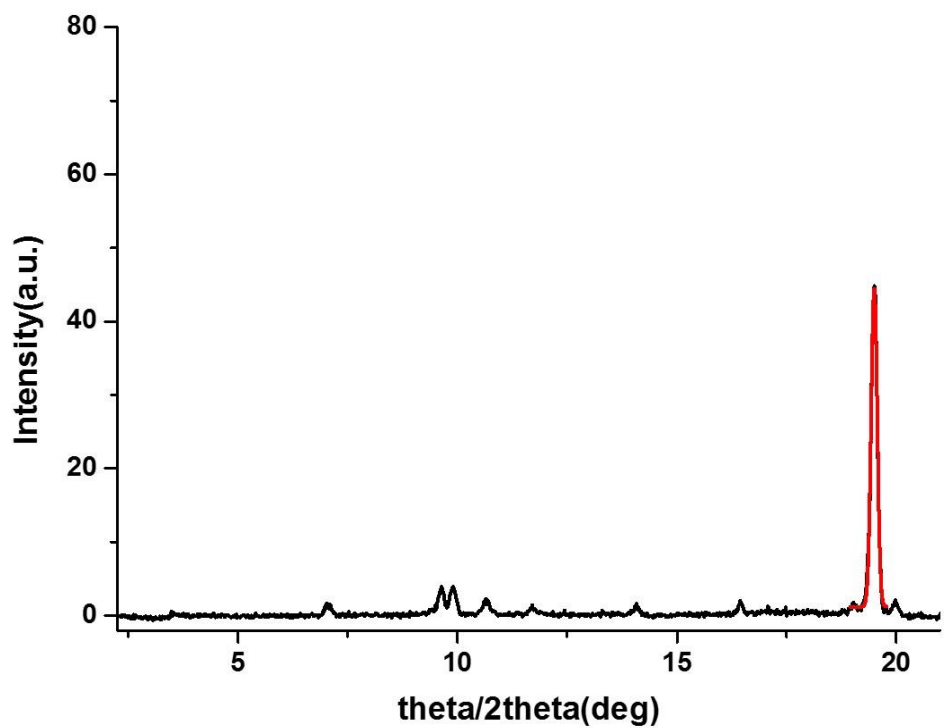

**Figure S12.** LT-XRD pattern measured at -40°C and fitting restuls for scherrer equation.

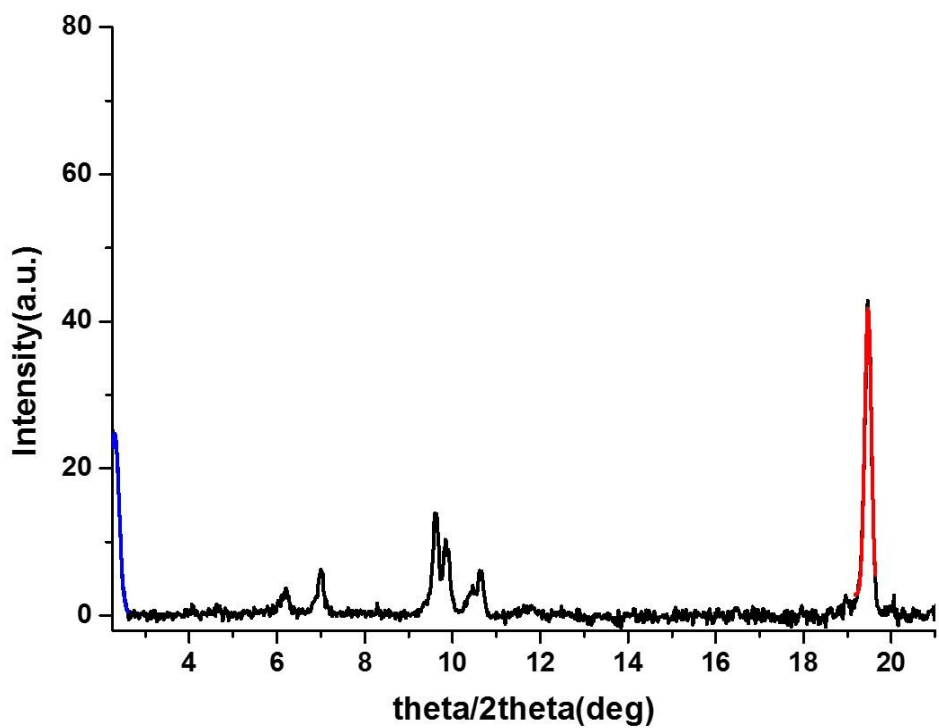

**Figure S13.** LT-XRD pattern measured at 12°C and fitting restuls for scherrer equation.

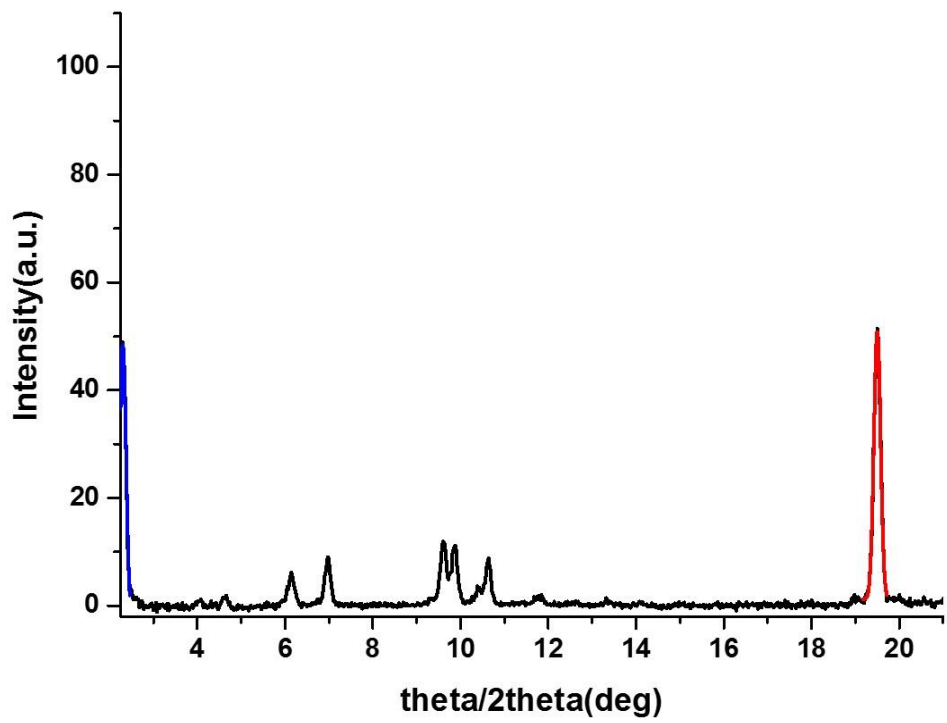

**Figure S14.** LT-XRD pattern measured at 18°C and fitting restuls for scherrer equation.

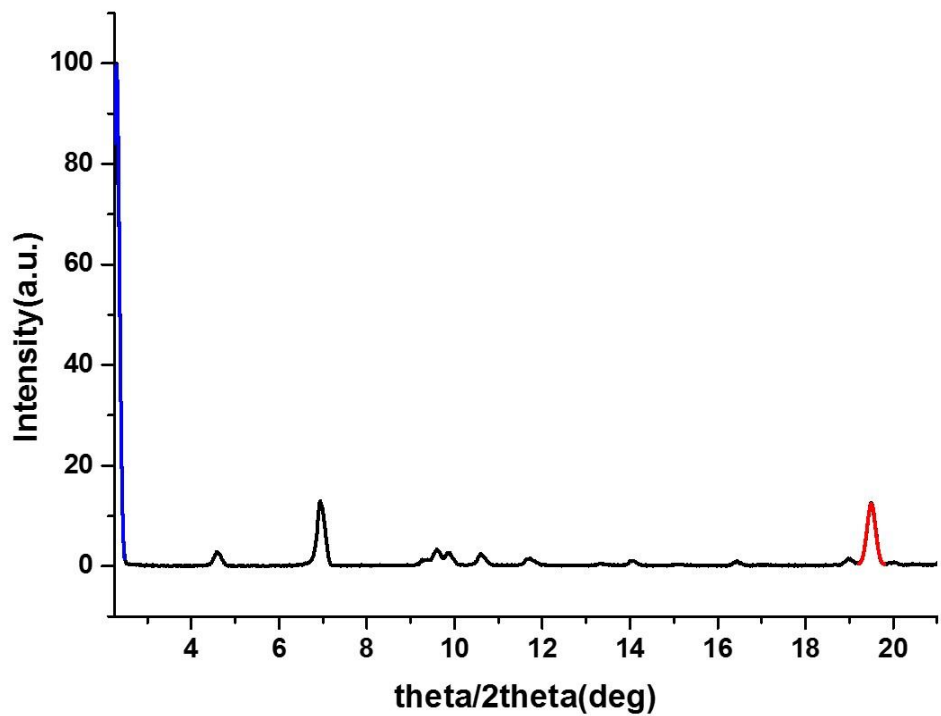

**Figure S15.** LT-XRD pattern measured at 28°C and fitting restuls for scherrer equation.

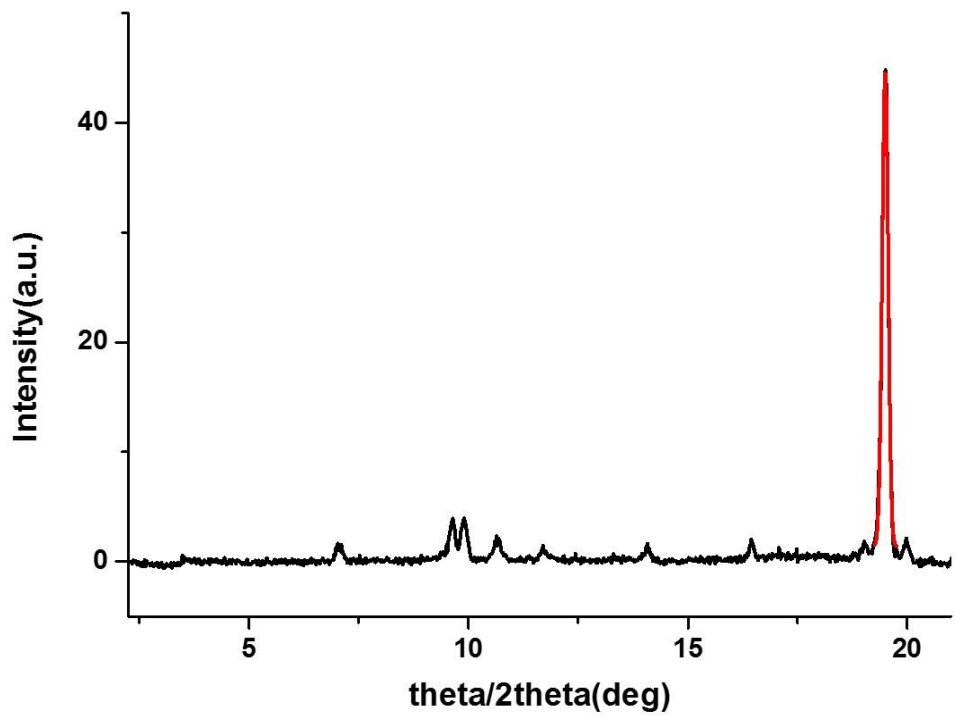

Figure S16. XRD pattern of as-synthezied third phase of HYLION-12 and fitting restuls for scherrer equation.

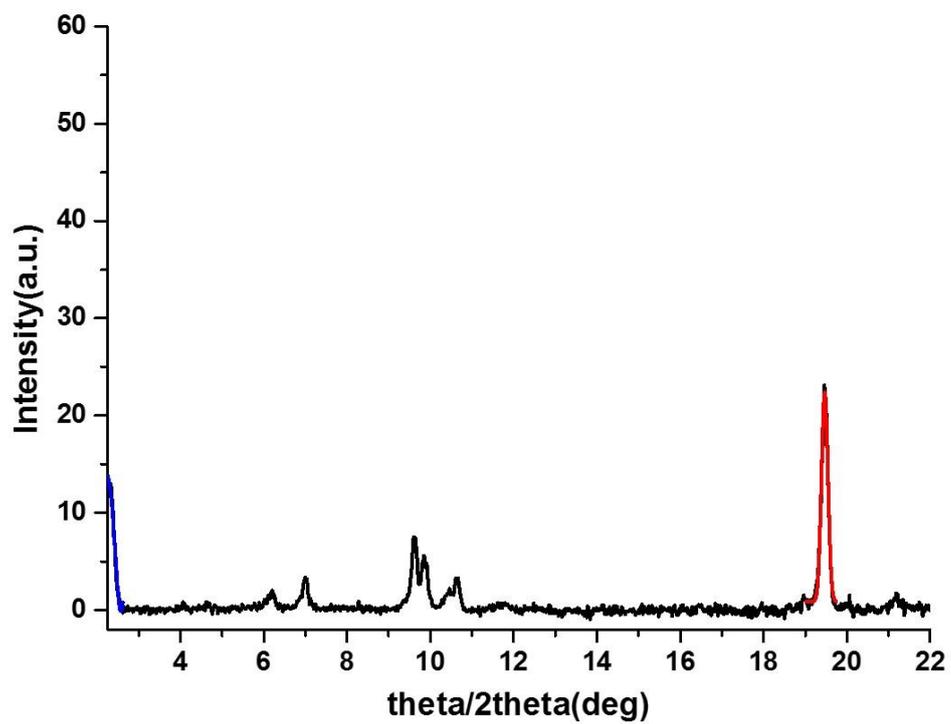

Figure S17. XRD pattern of third phase of HYLION-12 stored for 24h in saturated Ethyl acetate solution at -63°C and fitting restuls for scherrer equation.

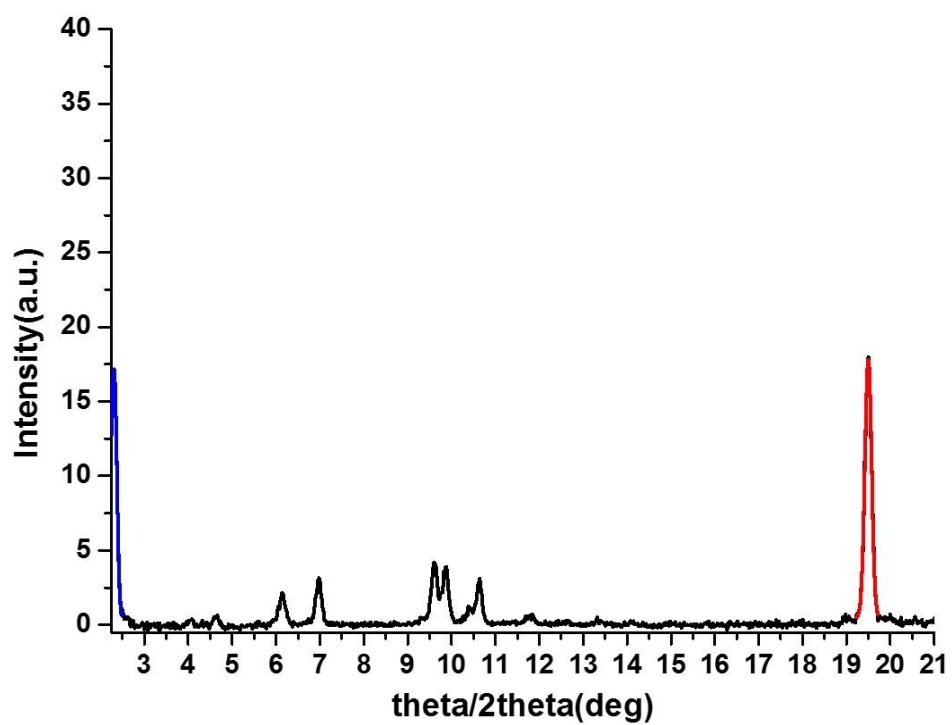

**Figure S18.** XRD pattern of third phase of HYLION-12 stored for 48h in saturated Ethyl acetate solution at -63°C and fitting restuls for scherrer equation.

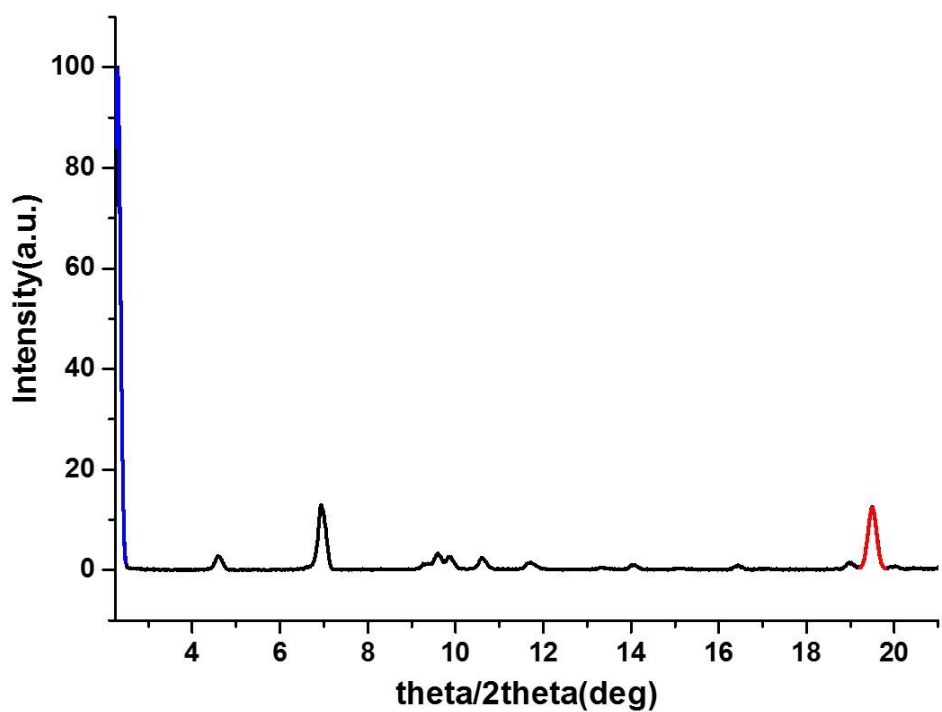

Figure S19. XRD pattern of third phase of HYLION-12 stored for 72h in saturated Ethyl acetate solution at -63°C and fitting restuls for scherrer equation.

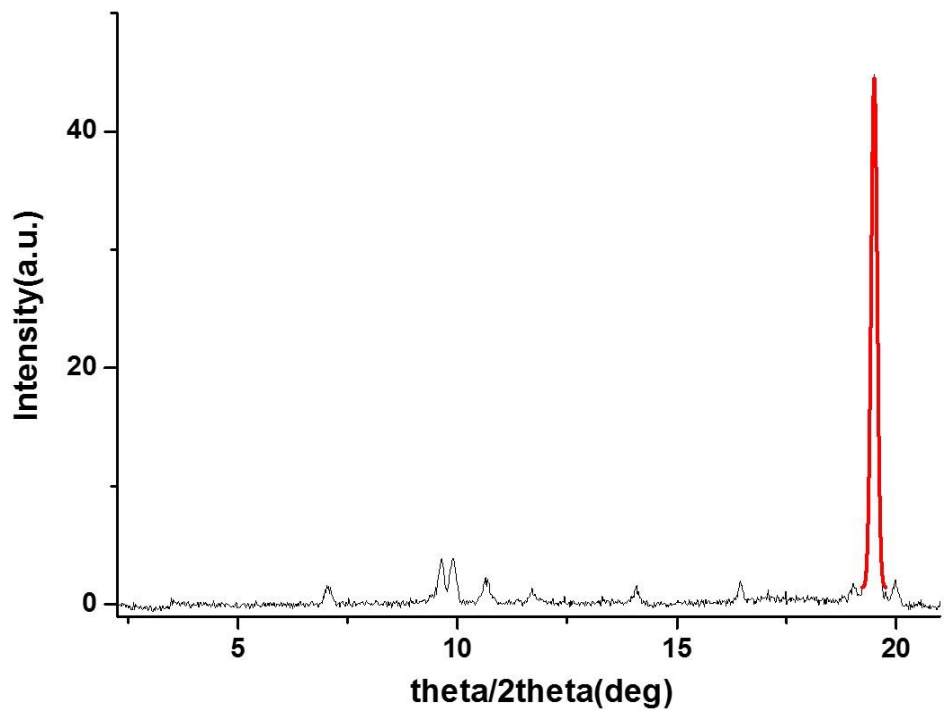

Figure S20. XRD pattern of third phase of HYLION-12 before pressure induced crysatllization and fitting restuls for scherrer equation.

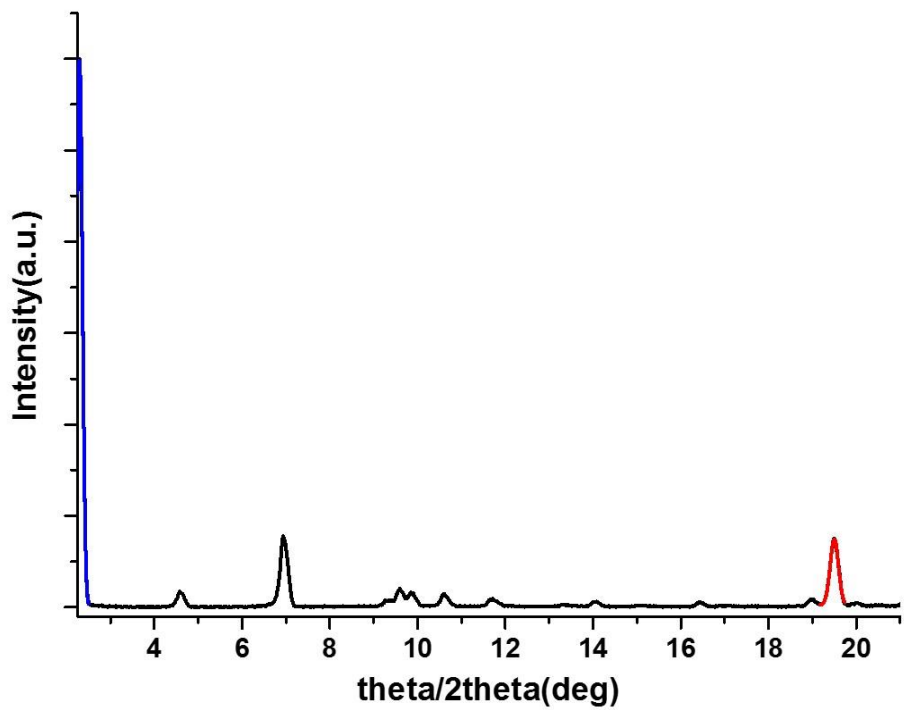

Figure S21. XRD pattern of third phase of HYLION-12 after 3.86 GPa pressure induced for crysatllization and fitting restuls for scherrer equation.

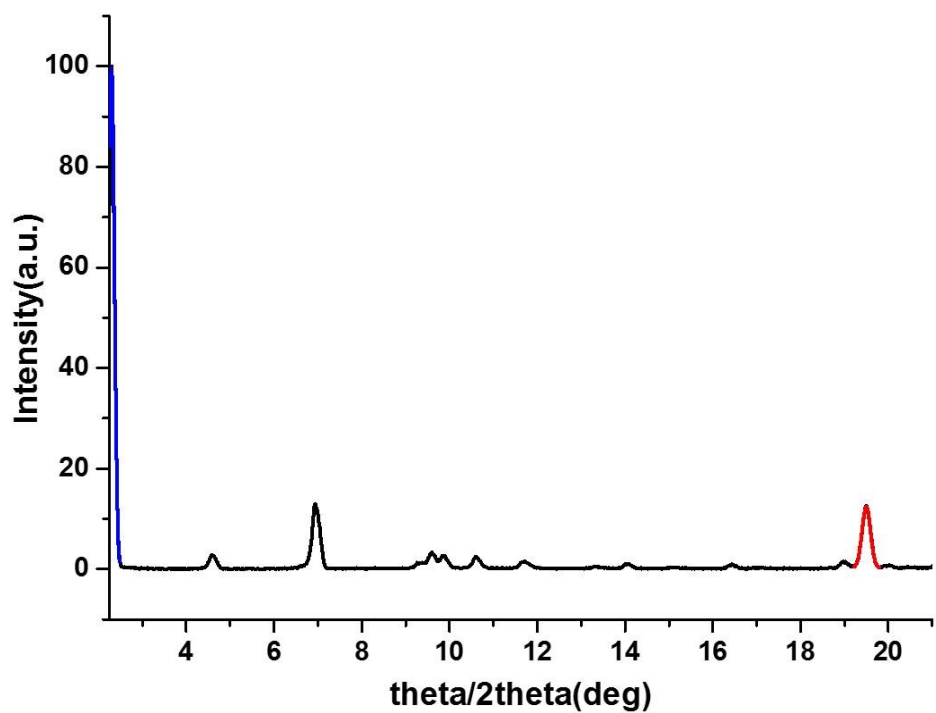

Figure S22. XRD pattern of third phase of HYLION-12 after 7.69 GPa pressure induced for crysatllization and fitting restuls for scherrer equation.

| Phase of HYLION-12 | Temperature (°C) | Grain size for xy-plane | Grain size for z-axis |
|---|---|---|---|
| $\Phi_n$ | -196 | 37.432 | 0 |
| $\Phi_n$ | -65 | 38.322 | 0 |
| $\Phi_n$ | -50 | 38.327 | 0 |
| $\Phi_o$ | -40 | 43.149 | 45.434 |
| $\Phi_o$ | -35 | 44.515 | 49.763 |
| $\Phi_o$ | -30 | 46.218 | 52.104 |
| $\Phi_o$ | -10 | 46.218 | 53.967 |
| $\Phi_o$ | -5 | 46.811 | 54.355 |
| $\Phi_o$ | 5 | 47.218 | 59.439 |
| $\Phi_o$ | 20 | 51.596 | 60.579 |

**Table S1. Grain size of HYLION-12 crysatllized with various temperature**

| Phase of HYLION-12 | Temperature (°C) | Grain size for xy-plane | Grain size for z-axis |
|---|---|---|---|
| $\Phi_n$ | -60 | 0 | 52.814 |
| $\Phi_n$ | -40 | 0 | 51.682 |
| $\Phi_o$ | 12 | 35.298 | 48.147 |
| $\Phi_o$ | 18 | 48.195 | 46.914 |
| $\Phi_o$ | 28 | 54.323 | 46.279 |

**Table S2. Grain size of HYLION-12 measured during LT-XRD measurment**

| Phase of HYLION-12 | Time (h) | Grain size for xy-plane | Grain size for z-axis |
|---|---|---|---|
| $\Phi_n$ | 0 | 0 | 51.651 |
| $\Phi_o$ | 24 | 25.133 | 46.174 |
| | 48 | 54.652 | 46.839 |
| | 72 | 57.023 | 46.472 |

**Table S3. Grain size of HYLION-12 measured during LT-XRD measurment**

| Phase of HYLION-12 | Pressure(GPa) | Grain size for xy-plane | Grain size for z-axis |
|---|---|---|---|
| $\Phi_n$ | 0 | 0 | 51.693 |
| $\Phi_o$ | 3.86 | 54.88 | 50.860 |
| | 7.96 | 58.48 | 51.852 |

**Table S4. Press inducced grain size of HYLION-12**